\documentclass{article}
\usepackage{waspaa19,amssymb,amsmath,graphicx,times,url}
\usepackage{color}
\usepackage{blindtext}
\usepackage{pgfplots,tikz}
\usepackage[subtle, mathspacing=normal, wordspacing=tight, bibbreaks=normal, floats=normal, tracking=normal, paragraphs=tight]{savetrees}
\usepackage{cite}

\newcommand{\mvec}[1]{{\mbox{\boldmath $#1$}}}
\newcommand{\mrm}[1]{\mathrm{#1}}

\title{RTF-steered binaural MVDR beamforming incorporating\\
multiple external microphones}

\name{Nico G\"o{\ss}ling, Wiebke Middelberg, Simon Doclo}
\address{University of Oldenburg, Department of Medical Physics and Acoustics\\
	and Cluster of Excellence Hearing4all, Oldenburg, Germany\\
	\texttt{nico.goessling@uni-oldenburg.de}
	\thanks{This work was funded by
the Deutsche Forschungsgemeinschaft (DFG, German Research Foundation) - Project ID 352015383 (SFB 1330 B2) and Project ID 390895286 (EXC 2177/1).}}

\begin{document}
\ninept
\maketitle

\begin{sloppy}

\begin{abstract}
  The binaural minimum-variance distortionless-response (BMVDR) beamformer is a well-known noise reduction algorithm that can be steered using the relative transfer function (RTF) vector of the desired speech source.
  Exploiting the availability of an external microphone that is spatially separated from the head-mounted microphones, an efficient method has been recently proposed to estimate the RTF vector in a diffuse noise field.
  When multiple external microphones are available, different RTF vector estimates can be obtained by using this method for each external microphone.
  In this paper, we propose several procedures to combine these RTF vector estimates, either by selecting the estimate corresponding to the highest input SNR, by averaging the estimates or by combining the estimates in order to maximize the output SNR of the BMVDR beamformer.
  Experimental results for a moving speaker and diffuse noise in a reverberant environment show that the output SNR-maximizing combination yields the largest binaural SNR improvement and also outperforms the state-of-the art covariance whitening method.
\end{abstract}


\section{Introduction}
\label{sec:intro}
Noise reduction algorithms for head-mounted assistive listening devices (e.g., hearing aids, earbuds, headsets) are crucial to improve speech intelligibility and speech quality in noisy environments.
Binaural noise reduction algorithms, which exploit the information captured by all microphones on both sides of the head \cite{Doclo2015,Doclo2018}, do not only allow to reduce unwanted sound sources but also allow to preserve the listener's spatial impression of the acoustic scene.
As a well-known example, the binaural minimum-variance distortionless-response (BMVDR) beamformer is able to preserve the binaural cues (i.e. the interaural time and level differences) of a desired speech source \cite{Doclo2015,Doclo2018,Cornelis2010}.
For a moving speech source in a reverberant environment, the BMVDR can be steered using the relative transfer functions (RTFs) \cite{Gannot2001}, which relate the acoustic transfer functions between the desired speech source and all microphones to the so-called reference microphones.

To improve the performance of (binaural) algorithms in terms of noise reduction and source localization accuracy, it has been proposed to use an external microphone in conjunction with the head-mounted microphones \cite{Bertrand2009,Szurley2016,Farmani2017,Yee2017J,Ali2018,Ali2018_iwaenc,Goessling2018_iwaenc_a,Goessling2019_icassp}.
For a diffuse noise field, an efficient RTF vector estimation method has been proposed in \cite{Goessling2018_iwaenc_a}, which exploits the spatial coherence (SC) properties of the noise field.
More specifically, the SC method assumes that the noise component in the external microphone signal is uncorrelated with the noise components in the head-mounted microphone signals.
%The SC method then yields an unbiased RTF vector estimate, except for a biased estimate of the RTF corresponding to the external microphone.
%In \cite{Goessling2019_icassp} however, it has been shown that this bias is real-valued (hence not affecting the phase of the RTF vector estimate) and that in practice the external microphone signal can also be included in the BMVDR processing.

In this paper, we consider the more general 	scenario with multiple external microphones.
Using the SC method, each external microphone yields a (different) RTF vector estimate, such that the question arises how to combine these RTF vector estimates.
In the first procedure, we propose to select the RTF vector estimate corresponding to the external microphone with the highest narrowband signal-to-noise ratio (SNR).
In the second procedure, we propose to simply average the different RTF vector estimates.
In the third procedure, we propose to linearly combine the different RTF vector estimates such that the narrowband output SNR of the BMVDR is maximized.
Experimental results of an on-line implementation of the BMVDR using recorded signals of a moving speaker and diffuse noise in a reverberant environment are provided.
The results show that the output SNR-maximizing combination of the SC-based RTF vector estimates leads to the largest binaural SNR improvement compared to the other procedures and the state-of-the-art covariance whitening method \cite{Markovich2009,Markovich2015}.
\section{Configuration and Notation}
\label{sec:config}
Consider the binaural hearing device configuration depicted in Figure \ref{fig:config}, consisting of a left and a right hearing device (each equipped with $M_\mrm{D}$ microphones), and $M_\mrm{E}$ external microphones that are spatially separated from the head-mounted microphones, i.e. $M=2M_\mrm{D} + M_\mrm{E}$ microphones in total.
In the frequency-domain, the $m$-th microphone signal of the left device can be written as
\begin{equation} \label{eq:Y}
	y_{\mrm{L},m}(\omega) = x_{\mrm{L},m}(\omega) + n_{\mrm{L},m}(\omega)\, , \quad m \in \{1,\dots,M_\mrm{D}\}\, ,
\end{equation}
with $x_{\mrm{L},m}(\omega)$ the desired speech component and $n_{\mrm{L},m}(\omega)$ the noise component.
For the sake of conciseness, the frequency $\omega$ will be omitted in the remainder of the paper.
The $m$-th microphone signal of the right device $y_{\mrm{R},m}$ and the $i$-th external microphone signal $y_{\mrm{E},i}$ are defined similarly as in \eqref{eq:Y}.
The $M$-dimensional microphone signal vector, containing \textit{all} microphone signals, is defined as
\begin{equation} \label{eq:y}
	\mvec{y} = \left[y_{\mrm{L},1},\,\dots,\,y_{\mrm{L},M_\mrm{D}},\,y_{\mrm{R},1},\,\dots,\,y_{\mrm{R},M_\mrm{D}},\,y_{\mrm{E},1},\,\dots,\,y_{\mrm{E},M_\mrm{E}} \right]^T \, ,
\end{equation}
with $(\cdot)^T$ denoting the transpose operator. Using \eqref{eq:Y}, the vector $\mvec{y}$ can be written as
\begin{equation}
	\mvec{y} = \mvec{x} + \mvec{n} \, ,
\end{equation}
where the speech vector $\mvec{x}$ and the noise vector $\mvec{n}$ are defined similarly as in \eqref{eq:y}.
Without loss of generality, the first microphone on each device is chosen as the reference microphone, i.e.
\begin{equation}
	y_\mrm{L} = y_{\mrm{L},1} = \mvec{e}_\mrm{L}^T\mvec{y}\, , \quad y_\mrm{R} = y_{\mrm{R},1} = \mvec{e}_\mrm{R}^T\mvec{y}\, ,
\end{equation}
where $\mvec{e}_\mrm{L}$ and $\mvec{e}_\mrm{R}$ denote selection vectors consisting of zeros and one element equal to 1.
Assuming a single desired speech source, the vector $\mvec{x}$ can be written as
\begin{equation}
	\mvec{x} = \mvec{a}_\mrm{L}x_\mrm{L} = \mvec{a}_\mrm{R}x_\mrm{R} \, ,
\end{equation}
where $\mvec{a}_\mrm{L}$ and $\mvec{a}_\mrm{R}$ denote the $M$-dimensional RTF vectors of the desired speech source with respect to the reference microphones on the left and the right device, respectively.
It should be noted that one of the elements of the RTF vectors (corresponding to the reference microphone) is equal to 1 and that the RTF vectors are related as $\mvec{a}_\mrm{R} = \mvec{a}_\mrm{L} / \mvec{e}_\mrm{R}^T\mvec{a}_\mrm{L}$.
The noisy input covariance matrix $\mvec{R}_\mrm{y}$, the speech covariance matrix $\mvec{R}_\mrm{x}$ and the noise covariance matrix $\mvec{R}_\mrm{n}$ are defined as
\begin{align}
	\mvec{R}_\mrm{y} =  \mathcal{E}\{\mvec{y}\mvec{y}^H\}\, , \;
	\mvec{R}_\mrm{x} =  \mathcal{E}\{\mvec{x}\mvec{x}^H\}\, , \;
	\mvec{R}_\mrm{n} = \mathcal{E}\{\mvec{n}\mvec{n}^H\} \, ,
\end{align}
where $\mathcal{E}\{\cdot\}$ denotes the expectation operator and $(\cdot)^H$ denotes the conjugate transpose.
Assuming statistical independence between the desired speech component and the noise component, the noisy input covariance matrix is equal to $\mvec{R}_\mrm{y} = \mvec{R}_\mrm{x} + \mvec{R}_\mrm{n}$.

The output signals of the left and right devices are calculated by filtering and summing \textit{all} microphone signals, i.e. the head-mounted microphone signals as well as the external microphone signals, using the complex-valued filter vectors $\mvec{w}_\mrm{L}$ and $\mvec{w}_\mrm{R}$ (see Figure \ref{fig:config}), i.e.
\begin{equation}
	z_\mrm{L} = \mvec{w}_\mrm{L}^H\mvec{y} \, ,  \quad z_\mrm{R} = \mvec{w}_\mrm{R}^H\mvec{y} \, .
\end{equation}
The input SNR of the $m$-th microphone signal is given by the ratio of the input power spectral density (PSD) of the desired speech component and the input PSD of the noise component, i.e.
\begin{equation} \label{eq:iSNR}
	\mrm{SNR}_m^\mrm{in} = \frac{\mvec{e}_{m}^T\mvec{R}_\mrm{x}\mvec{e}_{m}}{\mvec{e}_m^T\mvec{R}_\mrm{n}\mvec{e}_{m}} = \frac{\mvec{e}_{m}^T\mvec{R}_\mrm{y}\mvec{e}_{m}}{\mvec{e}_m^T\mvec{R}_\mrm{n}\mvec{e}_{m}} - 1 \,,
\end{equation}
with $\mvec{e}_{m}$ an $M$-dimensional vector selecting the element corresponding to the $m$-th microphone.
Similarly, the output SNR of the left and the right output signals is given by the ratio of the output PSD of the desired speech component and the output PSD of the noise component, i.e.
\begin{equation} \label{eq:outSNR}
	\mrm{SNR}_\mrm{L}^\mrm{out} = \frac{\mvec{w}_\mrm{L}^H\mvec{R}_\mrm{x}\mvec{w}_\mrm{L}}{\mvec{w}_\mrm{L}^H\mvec{R}_\mrm{n}\mvec{w}_\mrm{L}} \,, \quad \mrm{SNR}_\mrm{R}^\mrm{out} = \frac{\mvec{w}_\mrm{R}^H\mvec{R}_\mrm{x}\mvec{w}_\mrm{R}}{\mvec{w}_\mrm{R}^H\mvec{R}_\mrm{n}\mvec{w}_\mrm{R}} \, .
\end{equation}
\begin{figure}
	\centering
	\includegraphics{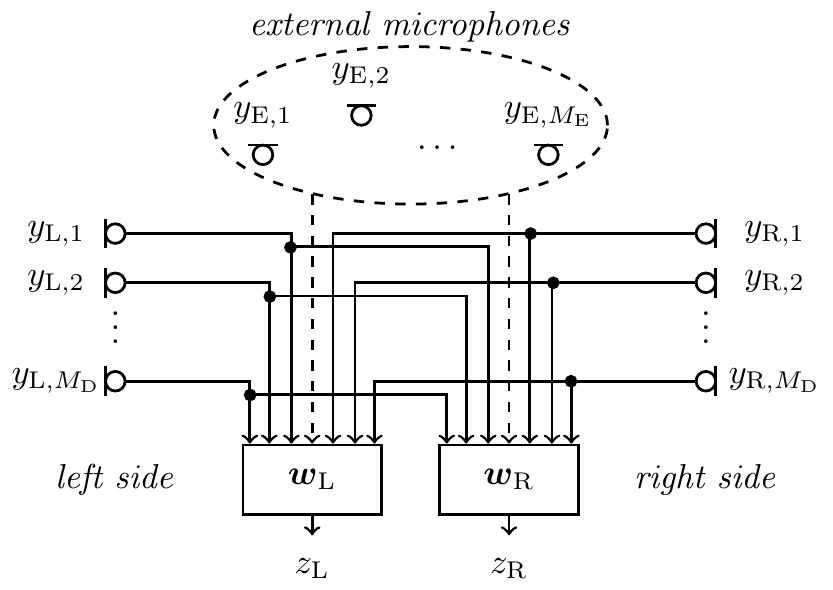}
	\vspace{-6mm}
	\caption{Binaural hearing device configuration incorporating multiple external microphones.}
	\label{fig:config}
\end{figure}
\section{Binaural MVDR Beamformer}
\label{sec:bmvdr}
The BMVDR \cite{Doclo2018,Doclo2010} aims at minimizing the output noise PSD while preserving the desired speech component in the reference microphone signals ($x_\mrm{L}$ and $x_\mrm{R}$), hence preserving the binaural cues of the desired speech source.
The optimization problem for the left filter vector $\mvec{w}_\mrm{L}$ is given by
\begin{equation} \label{eq:optiMVDR}
	\min_{\mvec{w}_\mrm{L}} \; \mvec{w}_\mrm{L}^H\mvec{R}_\mrm{n}\mvec{w}_\mrm{L} \quad \text{subject to} \quad \mvec{w}_\mrm{L}^H\mvec{a}_\mrm{L} = 1 \, .
\end{equation}
The optimization problem for the right filter vector $\mvec{w}_\mrm{R}$ is defined similarly.
The filter vectors solving the optimization problems are equal to \cite{Doclo2015,Doclo2018,Doclo2010}
\begin{equation} \label{eq:mvdr}
	\mvec{w}_\mrm{L} = \frac{\mvec{R}_\mrm{n}^{-1}\mvec{a}_\mrm{L}}{\mvec{a}_\mrm{L}^H\mvec{R}_\mrm{n}^{-1}\mvec{a}_\mrm{L}}\, , \quad \mvec{w}_\mrm{R} = \frac{\mvec{R}_\mrm{n}^{-1}\mvec{a}_\mrm{R}}{\mvec{a}_\mrm{R}^H\mvec{R}_\mrm{n}^{-1}\mvec{a}_\mrm{R}}\, .
\end{equation}
Hence, estimates of the noise covariance matrix $\mvec{R}_\mrm{n}$ and the RTF vectors $\mvec{a}_\mrm{L}$ and $\mvec{a}_\mrm{R}$ are required to compute the BMVDR filter vectors in practice.
Typically, the noise covariance matrix $\mvec{R}_\mrm{n}$ is recursively estimated from the microphone signals during speech pauses, e.g., based on a voice activity detector or speech presence probability \cite{Gerkmann2012}.

The following sections describe different methods to estimate the RTF vectors $\mvec{a}_\mrm{L}$ and $\mvec{a}_\mrm{R}$.
Section 4 describes the covariance whitening method, which is a state-of-the-art RTF vector estimation method for a general noise field.
In Section 5 we propose RTF vector estimation methods that assume that the noise component in each external microphone signal is uncorrelated with the noise components in all other microphone signals.
\section{Covariance Whitening Method} \label{sec:CW}
The covariance whitening (CW) method \cite{Markovich2009,Markovich2015} is based on the generalized eigenvalue decomposition of the noisy input covariance matrix $\mvec{R}_\mrm{y}$ and the noise covariance matrix $\mvec{R}_\mrm{n}$.
Using the Cholesky decomposition of the noise covariance matrix, i.e.
\begin{equation} \label{eq:cholesky}
	\mvec{R}_\mrm{n} = \mvec{R}_\mrm{n}^{H/2}\mvec{R}_\mrm{n}^{1/2} \, ,
\end{equation}
the pre-whitened noisy input covariance matrix is defined as
\begin{equation} \label{eq:pwRy}
	\mvec{R}_\mrm{y}^\mrm{w} = \mvec{R}_\mrm{n}^{-H/2}\mvec{R}_\mrm{y}\mvec{R}_\mrm{n}^{-1/2} \, .
\end{equation}
Using \eqref{eq:cholesky} and \eqref{eq:pwRy}, the left RTF vector can be estimated as \cite{Markovich2015}
\begin{equation} \label{eq:cw}
	\mvec{a}_{\mrm{L}}^{\mrm{CW}} = \frac{\mvec{R}_\mrm{n}^{1/2}\mvec{p}}{\mvec{e}^T_\mrm{L}\mvec{R}_\mrm{n}^{1/2}\mvec{p}} \, ,
\end{equation}
with $\mvec{p} = \mathcal{P}\{\mvec{R}_\mrm{y}^\mrm{w}\}$ the principal eigenvector (corresponding to the largest eigenvalue) of the pre-whitened noisy input covariance matrix $\mvec{R}_\mrm{y}^\mrm{w}$.
Due to the Cholesky decomposition and the $M \times M$-dimensional eigenvalue decomposition (EVD) the CW method typically has a rather large computational complexity, especially for large  $M$.
\section{Spatial Coherence Method}
\label{sec:rtf}
In this section, we propose RTF vector estimation methods that assume that the noise component in each external microphone signal is uncorrelated with the noise components in all other microphone signals.
This can, e.g., be assumed for a diffuse noise field when the external microphones are spatially separated from each other and from the head-mounted microphones.
In Section 5.1, we review the SC method as presented in \cite{Goessling2018_iwaenc_a} for one external microphone.
In Section 5.2, we propose three different procedures to linearly combine the RTF vector estimates obtained by using the SC method for each external microphone.
\subsection{SC method per external microphone}
\label{sec:scOne}
If the noise component in the $i$-th external microphone signal is uncorrelated with the noise components in all other microphone signals, it has been shown in \cite{Goessling2018_iwaenc_a,Goessling2019_icassp} that the left RTF vector can be efficiently estimated from the noisy input covariance matrix $\mvec{R}_\mrm{y}$ using the SC method as
\begin{equation} \label{eq:sc1}
	\mvec{a}_{\mrm{L}}^{\mrm{SC-}i} = \frac{\mvec{R}_\mrm{y}\mvec{e}_{\mrm{E},i}}{\mvec{e}_\mrm{L}^T\mvec{R}_\mrm{y}\mvec{e}_{\mrm{E},i}} \, , \quad i \in \{1,\dots,M_\mrm{E}\} \, ,
\end{equation}
with $\mvec{e}_{\mrm{E},i}$ an $M$-dimensional vector, selecting the element corresponding to the $i$-th external microphone.
The estimator in \eqref{eq:sc1} yields an unbiased RTF vector estimate, except for a biased estimate of the RTF corresponding to the $i$-th external microphone.
However, in \cite{Goessling2019_icassp} it has been shown that this bias is real-valued (hence not affecting the phase of the RTF vector estimate), depends on the input SNR in the $i$-th external microphone and typically can be neglected in practice.
\subsection{Combination of SC-based RTF vector estimates}
\label{sec:scMult}
Since in practice an estimate of the noisy input covariance matrix $\hat{\mvec{R}}_\mrm{y}$ is used in \eqref{eq:sc1}, typically $M_\mrm{E}$ different SC-based RTF vector estimates are obtained, such that the question arises how to use these estimates.
In this paper, we propose to linearly combine the different RTF vector estimates (per frequency) and to use the resulting RTF vector in the BMVDR.
The (normalized) combined RTF vector estimate is given by
\begin{equation} \label{eq:SM}
	\boxed{
	\mvec{a}_\mrm{L}^\mrm{SC-C} = \frac{\mvec{A}^\mrm{SC}_\mrm{L}\mvec{c}}{\mvec{e}^T_\mrm{L}\mvec{A}^\mrm{SC}_\mrm{L}\mvec{c}}
		}
\end{equation}
with $\mvec{A}^\mrm{SC}_\mrm{L}$ an $M \times M_\mrm{E}$-dimensional matrix, containing the $M_\mrm{E}$ SC-based RTF vector estimates, i.e.
\begin{equation} \label{eq:SCmatrix}
	\mvec{A}^\mrm{SC}_\mrm{L} = \left[ \mvec{a}_{\mrm{L}}^{\mrm{SC-}1}, \dots,  \mvec{a}_{\mrm{L}}^{\mrm{SC-}M_\mrm{E}}\right] \, ,
\end{equation}
and $\mvec{c}$ an $M_\mrm{E}$-dimensional (complex-valued) combination vector.
Please note that the combination closest to the true RTF vector $\mvec{a}_\mrm{L}$ could be obtained by orthogonally projecting $\mvec{a}_\mrm{L}$ on the column space of $\mvec{A}^\mrm{SC}_\mrm{L}$, which is obviously not possible in practice.
In the following, we hence propose three different procedures to determine the combination vector $\mvec{c}$ in practice.

The first procedure, denoted as \textbf{iSNR}, is to select the RTF vector estimate (per frequency) corresponding to the external microphone with the highest narrowband input SNR, similarly to \cite{Lawin-Ore2012}.
Due to \eqref{eq:iSNR}, this only requires an estimate of $\mvec{R}_\mrm{y}$ (and not $\mvec{R}_\mrm{x}$), i.e.
\begin{equation} \label{eq:bestSNR}
	\mvec{c}^\mrm{iSNR} = \mvec{e}_{\mrm{E},\hat{i}} \, , \quad
	\hat{i} = \arg\max_i \; \frac{\mvec{e}_{\mrm{E},i}^T\mvec{R}_\mrm{y}\mvec{e}_{\mrm{E},i}}{\mvec{e}_{\mrm{E},i}^T\mvec{R}_\mrm{n}\mvec{e}_{\mrm{E},i}} \, .
\end{equation}
Especially for a dynamic acoustic scenario with a moving speaker, the iSNR-based selection procedure is expected to outperform the SC method only using on one external microphone.

Assuming a uniform distribution of the estimation errors for the SC-based RTF vector estimates, in the second procedure, denoted as \textbf{AV}, we propose to simply average the estimates, i.e.
\begin{equation} \label{eq:av}
		\mvec{c}^{\mrm{AV}} = \left[\frac{1}{M_\mrm{E}}, \dots, \frac{1}{M_\mrm{E}} \right]^T \, .
\end{equation}
Intuitively, this procedure is sub-optimal, especially when the estimation errors are very different. 
%which probably is a sub-optimal use of the available information.

As a more sophisticated procedure, denoted as \textbf{mSNR}, we propose to combine the SC-based RTF vector estimates (per frequency) such that the narrowband output SNR of the BMVDR is maximized.
Using \eqref{eq:SM} in \eqref{eq:mvdr}, the left output SNR in \eqref{eq:outSNR} can be written as the generalized Rayleigh quotient
\begin{equation} \label{eq:mvdrSNR}
	\mrm{SNR}_\mrm{BMVDR,L}^\mrm{out	} = \frac{\mvec{c}^H \mvec{\Lambda}_1 \mvec{c}}{\mvec{c}^H\mvec{\Lambda}_2 \mvec{c}} - 1\,,
\end{equation}
with
\begin{align}
	\mvec{\Lambda}_1 &= (\mvec{A}^\mrm{SC}_\mrm{L})^H\mvec{R}_\mrm{n}^{-1}\mvec{R}_\mrm{y}\mvec{R}_\mrm{n}^{-1} \mvec{A}^\mrm{SC}_\mrm{L} \, ,\\
	\mvec{\Lambda}_2 &= (\mvec{A}^\mrm{SC}_\mrm{L})^H\mvec{R}_\mrm{n}^{-1}\mvec{A}^\mrm{SC}_\mrm{L} \, .
\end{align}
Aiming at maximizing the output SNR of the BMVDR, the SNR-maximizing combination vector $\mvec{c}^\mrm{mSNR}$ is equal to the principal eigenvector of the $M_\mrm{E} \times M_\mrm{E}$-dimensional matrix $\mvec{\Lambda}_2^{-1}\mvec{\Lambda}_1$, i.e.
\begin{equation} \label{eq:maxalpha}
	\boxed{
	\mvec{c}^\mrm{mSNR} = \arg\max_\mvec{c}\; \mrm{SNR}_\mrm{BMVDR,L}^\mrm{out} = \mathcal{P}\{\mvec{\Lambda}_2^{-1}\mvec{\Lambda}_1\}
	}
\end{equation}
which hence also only requires an estimate of $\mvec{R}_\mrm{y}$ (and not $\mvec{R}_\mrm{x}$).
Although constructing the matrices $\mvec{\Lambda}_1$ and $\mvec{\Lambda}_2$ comes with some computational complexity, the computational complexity of the $M_\mrm{E} \times M_\mrm{E}$-dimensional EVD is always smaller than the $M \times M$-dimensional EVD required for the CW method (cf. Section \ref{sec:CW}).
\vspace{-0.68mm}
\section{Experimental Results}
\label{sec:experimental}
For a dynamic acoustic scenario with a moving speaker in a reverberant room, in this section we compare the performance of the BMVDR using the different RTF vector estimation methods described in Sections 4 and 5 for a binaural hearing device incorporating three external microphones.
\subsection{Recording setup and implementation}
\label{sec:rec}
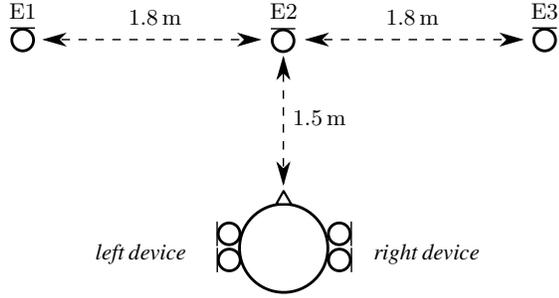
\begin{figure}
	\centering
	\begin{tikzpicture}[scale=0.95, transform shape]
		\node at (0,0) {\includegraphics[width=.9\linewidth]{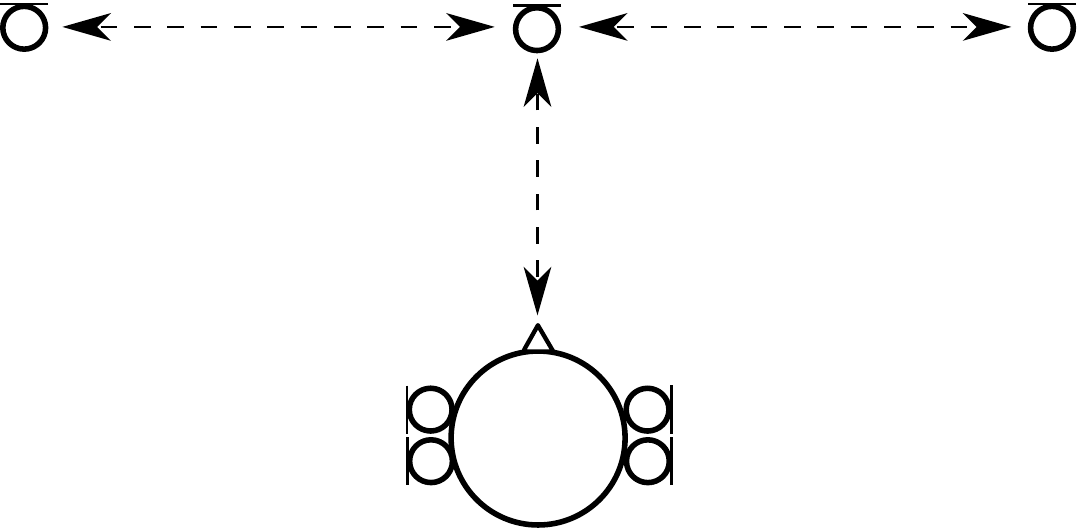}};
		\node at (-2,-1.3) {{\it left device}};
		\node at (2,-1.3) {{\it right device}};
		\node at (-3.65,2.1) {$\mrm{E}1$};
		\node at (3.65,2.1) {$\mrm{E}3$};
		\node at (0,2.1) {$\mrm{E}2$};
		\node at (0.5,0.6) {$1.5 \, \mrm{m}$};
		\node at (-1.8,2) {$1.8 \, \mrm{m}$};
		\node at (1.8,2) {$1.8 \, \mrm{m}$};
	\end{tikzpicture}
	\vspace{-3mm}
	\caption{Experimental setup with BTE hearing devices mounted on a dummy head and three external microphones.}
	\label{fig:setup}
\end{figure}
All signals were recorded in a laboratory where the reverberation time can be varied using absorber panels mounted on the walls and the ceiling.
The room dimensions are about $(7 \times 6 \times 2.7) \, \mrm{m}^3$ and the reverberation time was set to approximately $400 \, \mrm{ms}$.
A KEMAR dummy head was placed approximately in the center of the room with two behind-the-ear (BTE) hearing devices mounted to the ears.
Two microphones per hearing device, i.e. $M_\mrm{D} = 2$, with an inter-microphone distance of about $7 \, \mrm{mm}$ were used.
In addition, $M_\mrm{E}=3$ external microphones were placed in front of the dummy head as depicted in Figure \ref{fig:setup}.
Hence, in total $M=7$ microphones were used for the BMVDR.
The desired speech source was a male speaker, walking from the first external microphone ($\mrm{E}1$) to the third external microphone ($\mrm{E}3$) while speaking ten German sentences with pauses of about half a second between the sentences.
Pseudo-diffuse background noise was generated using four loudspeakers facing the corners of the laboratory, playing back different multi-talker recordings.
The desired speech source and the background noise were recorded separately and mixed afterwards.
Due to the moving speaker, the input SNR in the head-mounted reference microphone signals varied between approximately 0 and 6 dB, while the input SNR in the external microphone signals varied approximately between 0 and 11 dB.
All signals were recorded synchronously, hence neglecting synchronization and latency aspects.

All signals were sampled at a sampling rate of 16 kHz and processed in the short-time Fourier transform domain using a 32 ms square-root Hann window with 50\% overlap.
To distinguish between speech-plus-noise and noise-only time-frequency bins, the estimated speech presence probabilities \cite{Gerkmann2012} in the three (noisy) external microphone signals were averaged and thresholded.
The noisy input covariance matrix $\mvec{R}_\mrm{y}$ and the noise covariance matrix $\mvec{R}_\mrm{n}$ were then recursively estimated during detected speech-plus-noise and noise-only bins, respectively, using time constants of 250 ms ($\mvec{R}_\mrm{y}$) and 1.5 s ($\mvec{R}_\mrm{n}$).

As performance measure, we used the binaural SNR improvement ($\Delta\mrm{BSNR}$), which is defined similarly as in \eqref{eq:iSNR} and \eqref{eq:outSNR} as
\begin{align}
	\Delta \mrm{BSNR} = &10 \log_{10}\left( \frac{\mvec{w}_\mrm{L}^H\mvec{R}_\mrm{x}\mvec{w}_\mrm{L} + \mvec{w}_\mrm{R}^H\mvec{R}_\mrm{x}\mvec{w}_\mrm{R}}{\mvec{w}_\mrm{L}^H\mvec{R}_\mrm{n}\mvec{w}_\mrm{L} + \mvec{w}_\mrm{R}^H\mvec{R}_\mrm{n}\mvec{w}_\mrm{R}} \right)\\ \nonumber
	 &- 10\log_{10}\left( \frac{\mvec{e}_\mrm{L}^H\mvec{R}_\mrm{x}\mvec{e}_\mrm{L} + \mvec{e}_\mrm{R}^H\mvec{R}_\mrm{x}\mvec{e}_\mrm{R}}{\mvec{e}_\mrm{L}^H\mvec{R}_\mrm{n}\mvec{e}_\mrm{L} + \mvec{e}_\mrm{R}^H\mvec{R}_\mrm{n}\mvec{e}_\mrm{R}} \right) \, .
\end{align}
The binaural SNR improvement was computed in the time-domain using the shadow filter approach.

Seven different RTF vector estimates were considered for the BMVDR in \eqref{eq:mvdr}:
\begin{itemize}
	\item The state-of-the-art \textbf{CW} estimate in \eqref{eq:cw}
	\item The SC estimate in \eqref{eq:sc1} using each external microphone separately, i.e. \textbf{SC-1}, \textbf{SC-2} and \textbf{SC-3}
	\item The proposed SC-C method using the combination vectors in \eqref{eq:bestSNR}, \eqref{eq:av} and \eqref{eq:maxalpha}, i.e. \textbf{iSNR}, \textbf{AV} and \textbf{mSNR}
\end{itemize}
\subsection{Results}
\label{sec:results}
\begin{figure}
	\centering
	\includegraphics{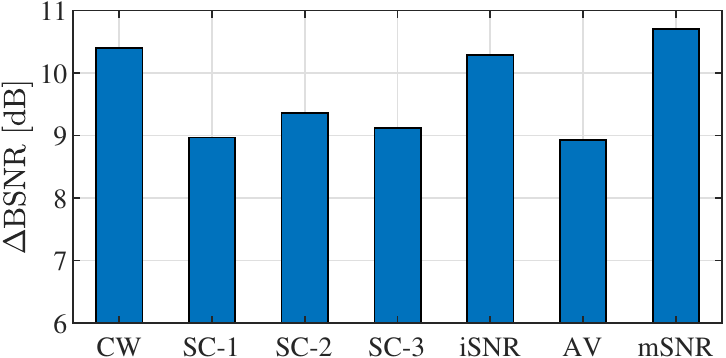}
	\vspace{-3mm}
	\caption{Binaural SNR improvement for all considered RTF vector estimation methods, averaged over time and frequency.}
	\label{fig:BSNR}
\end{figure}
\begin{figure}
	\centering
	\includegraphics{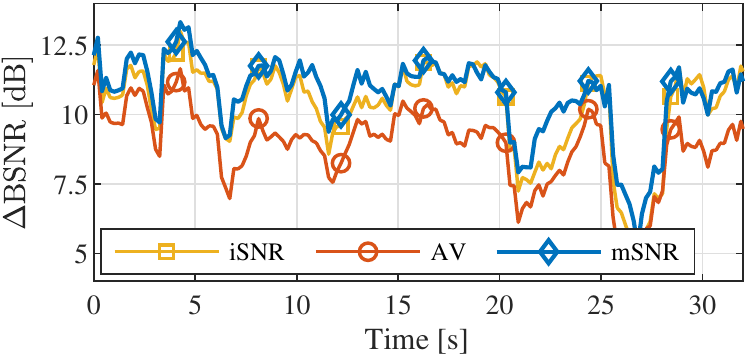}
	\vspace{-3mm}
	\caption{Binaural SNR improvement over time for the iSNR, AV and mSNR combination procedures, averaged over frequency.}
	\label{fig:BSNRTime}
\end{figure}
Figure \ref{fig:BSNR} depicts the $\Delta\mrm{BSNR}$ (averaged over time and frequency) for all considered RTF vector estimates.
The CW method as a state-of-the-art benchmark yields an average $\Delta\mrm{BSNR}$ of 10.4 dB.
The SC method using one external microphone, i.e. SC-1, SC-2 and SC-3, yields an average $\Delta\mrm{BSNR}$ of about 9 dB and hence could not reach the performance of the CW method.
The input SNR-based combination (iSNR) yields an average $\Delta\mrm{BSNR}$ of 10.3 dB, which is similar to the CW method.
The averaging combination (AV) yields an average $\Delta\mrm{BSNR}$ of only 8.9 dB, which is even worse than the SC method per external microphone.
This can probably be explained by the rather different RTF vector estimation errors for the three external microphones.
The SNR-maximizing combination (mSNR) yields an average $\Delta\mrm{BSNR}$ of 10.7 dB, hence outperforming all other combination procedures and RTF vector estimation methods.
Comparing the computational complexity of the best three methods, the CW method has the largest complexity due to the 7-dimensional EVD, whereas the mSNR combination only requires a 3-dimensional EVD and the iSNR combination does not even require an EVD.
Nevertheless, the mSNR combination enables to improve the $\Delta\mrm{BSNR}$ by about 0.5 dB compared to the iSNR combination.
Figure \ref{fig:BSNRTime} depicts the $\Delta\mrm{BSNR}$ over time (averaged over frequency) for the SC-C method using the proposed combination vectors in more detail.
It can be observed that the mSNR combination outperforms the iSNR and AV combination for almost all time instances.
The sound files of the input and output signals are available at \cite{AudioDemos}.
\section{Conclusions}
\label{sec:conclusions}
In this paper, we proposed to use the SC-based RTF vector estimation method for a scenario where multiple external microphones are incorporated into the BMVDR processing of a binaural hearing device.
Each external microphone was used to obtain an SC-based RTF vector estimate.
We proposed to linearly combine the different RTF vector estimates using an input SNR-based selection, simple averaging and a combination that maximizes the narrowband output SNR of the BMVDR.
Experimental evaluation in a dynamic scenario with a moving speaker in a reverberant environment showed that the SNR-maximizing combination yields the largest binaural SNR improvement and also outperforms the state-of-the art covariance whitening method.

%
% -------------------------------------------------------------------------
% Either list references using the bibliography style file IEEEtran.bst
\bibliographystyle{IEEEtran}
\bibliography{/Users/nico/Documents/Library/nicolib.bib}
%
% or list them by yourself
% \begin{thebibliography}{9}
% 
% \bibitem{waspaa19web}
%   \url{http://www.waspaa.com}.
%
% \bibitem{IEEEPDFSpec}
%   {PDF} specification for {IEEE} {X}plore$^{\textregistered}$,
%   \url{http://www.ieee.org/portal/cms_docs/pubs/confstandards/pdfs/IEEE-PDF-SpecV401.pdf}.
%
% \bibitem{PDFOpenSourceTools}
%   Creating high resolution {PDF} files for book production with 
%   open source tools, 
%   \url{http://www.grassbook.org/neteler/highres_pdf.html}.
%
% \bibitem{eWilliams1999}
% E. Williams, \emph{Fourier Acoustics: Sound Radiation and Nearfield Acoustic
%   Holography}. London, UK: Academic Press, 1999.
% 
% \bibitem{ieeecopyright}
%   \url{http://www.ieee.org/web/publications/rights/copyrightmain.html}.
%
% \bibitem{cJones2003}
% C. Jones, A. Smith, and E. Roberts, ``A sample paper in conference
%   proceedings,'' in \emph{Proc. IEEE ICASSP}, vol. II, 2003, pp. 803--806.
% 
% \bibitem{aSmith2000}
% A. Smith, C. Jones, and E. Roberts, ``A sample paper in journals,'' 
%   \emph{IEEE Trans. Signal Process.}, vol. 62, pp. 291--294, Jan. 2000.
% 
% \end{thebibliography}

\end{sloppy}
\end{document}